\documentclass[%
aip,amsmath,amssymb,reprint,%
]{revtex4-1}
\newcommand{\im}{\mathrm{i}}
\usepackage{graphicx}% Include figure files
\usepackage{dcolumn}% Align table columns on decimal point
\usepackage{bm}% 
\usepackage[utf8]{inputenc}
\usepackage[T1]{fontenc}
\usepackage{mathptmx}
\usepackage{etoolbox}
\usepackage{amsmath}

\makeatletter
\def\@email#1#2{%
	\endgroup
	\patchcmd{\titleblock@produce}
	{\frontmatter@RRAPformat}
	{\frontmatter@RRAPformat{\produce@RRAP{*#1\href{mailto:#2}{#2}}}\frontmatter@RRAPformat}
	{}{}
}%
\makeatother

\begin{document}
	
\preprint{AIP/123-QED}

\title{The Sakaguchi-Kuramoto model in presence of asymmetric interactions that break
phase-shift symmetry}

\author{M. Manoranjani}

\altaffiliation{ Centre for Nonlinear Science and Engineering, School of Electrical and Electronics Engineering, SASTRA Deemed University, Thanjavur 613 401, India}
\author{Shamik Gupta}
\altaffiliation{ Department of Physics, Ramakrishna Mission Vivekananda
	Educational and Research Institute, Belur Math, Howrah 711202, India}
\altaffiliation{Quantitative Life Sciences Section, ICTP - The Abdus Salam International Centre for Theoretical Physics, Strada Costiera 11, 34151 Trieste, Italy\\} 
\author{ V. K. Chandrasekar}
\altaffiliation{
	Centre for Nonlinear Science and Engineering, School of Electrical and Electronics Engineering, SASTRA Deemed University, Thanjavur 613 401, India}

\email{chandru25nld@gmail.com}

%\date{\today}

\date{\today}

\begin{abstract}
	The celebrated Kuramoto model provides an analytically tractable
	framework to study spontaneous collective synchronization and
	comprises globally coupled limit-cycle oscillators interacting
	symmetrically with one another. The Sakaguchi-Kuramoto model is a generalization of the
	basic model that considers the presence of a phase-lag
	parameter in the interaction, thereby making it asymmetric between
	oscillator pairs. Here, we consider a further generalization, by adding an interaction that breaks the phase-shift symmetry of the model. The highlight of our study is the
	unveiling of a very rich bifurcation diagram comprising of
	both oscillatory and non-oscillatory synchronized states as well as
	an incoherent state: There are regions of
	two-state as well as an interesting and hitherto unexplored three-state coexistence
	arising from asymmetric interactions in our model. 
\end{abstract}
\maketitle
\begin{quotation}
The Kuramoto model serves as a paradigm to study spontaneous synchronization, the phenomenon of synchronization among the phases of a macroscopic number of interacting limit-cycle oscillators of distributed frequencies. In the original model, the interaction is invariant with respect to arbitrary but equal shift of all the phases in the system (phase-shift symmetry), leading to either an unsynchronized/incoherent state or a synchronized state at long times. A previous study that considered explicit breaking of the phase-shift symmetry in the Kuramoto model revealed emergence of several interesting states such as an oscillatory (OS) state and a synchronized oscillatory death (OD) state, in addition to the incoherent (IC) state, so that phase-shift-symmetry-breaking interaction may be concluded to affect significantly the bifurcation diagram of the original Kuramoto model. In this work, we consider further implications of such symmetry-breaking interactions in a variant of the Kuramoto model, the so-called Sakaguchi-Kuramoto model in which the interaction between any pair of oscillators has unlike the Kuramoto model an explicit asymmetry.  Our model is a variant of the celebrated Winfree model of coupled oscillators.  Introducing a phase-shift-symmetry-breaking interaction in the Sakaguchi-Kuramoto model is found to significantly modify its bifurcation diagram and in particular result in the emergence of a very peculiar three-state-coexistence region between the OD-OS-IC, which we believe is new to the literature on Kuramoto and Sakaguchi-Kuramoto models. Our results are based on numerical integration of the dynamical equations as well as an exact analysis of the dynamics by invoking the so-called Ott-Antonsen ansatz that allows to derive a reduced set of time-evolution equations for the order parameter.  
\end{quotation}

%%%%%%%%%%%%%%%%%%%%%%%%%%%%%%%%%%%%%%%%%%%%%%%%%%%%%%%%%%%%%%%%%%%%%%%%%%%%%%%%%%%%%%%%%%%%
\section{Introduction}
\label{sec:intro}

The fundamental paradigm of the Kuramoto
model~\cite{Kuramoto:1984,Strogatz:2000,Acebron:2005,Gupta:2014,Gupta:2018}
has been widely employed over the years in studying collective behavior
of interacting stable limit-cycle oscillators. The model has been particularly
successful in explaining spontaneous
collective synchronization, a phenomenon exhibited by large ensembles of coupled
oscillators and encountered across disciplines, e.g., in
physics, biology, chemistry, ecology, electrical engineering,
neuroscience, and sociology~\cite{Pikovsky:2001}. Examples of collective synchrony include synchronized firing
of cardiac pacemaker cells~\cite{Peskin:1975}, synchronous emission of light
pulses by groups of fireflies~\cite{Buck:1988}, chirping of crickets~\cite{Walker:1969}, synchronization in
ensembles of electrochemical oscillators~\cite{Kiss:2002},
synchronization in human cerebral connectome~\cite{Schmidt:2015},  and
synchronous clapping of audience~\cite{Neda:2000}. The Kuramoto model
comprises $N$ globally-coupled stable limit-cycle oscillators with
distributed natural frequencies interacting symmetrically with one another
through the sine of their phase differences. Denoting by $\theta_j \in
[-\pi, \pi]$ the phase of the $j$-th oscillator, $j=1,2,\ldots,N$, the dynamics of the model is described by a set of $N$ coupled first-order nonlinear differential equations of the form
\begin{equation}
\frac{{\rm d}\theta_j}{{\rm d}t}=\omega_j+\frac{K}{N}\sum_{k=1}^N \sin(\theta_k-\theta_j),
\label{eq:eom-0}
\end{equation}
where $K \ge 0$ is the coupling constant, and $\omega_j$ is the natural
frequency of the $j$-th oscillator. The frequencies
$\{\omega_j\}$ denote a set of quenched-disordered random variables sampled
independently from a distribution $g(\omega)$ usually taken to be
unimodal, that is, symmetric about the mean $\omega_0$ and decreasing
monotonically and continuously to zero with increasing
$|\omega-\omega_0|$. In the dynamics~(\ref{eq:eom-0}) the interaction is
symmetric: the effect on the $j$-th oscillator due to
the $k$-th one being proportional to $\sin(\theta_k-\theta_j)$ is equal
in magnitude (but opposite in sign) to the one on the $k$-th
oscillator due to the $j$-th one. 

In contrast to the dynamics~(\ref{eq:eom-0}), interactions between
oscillators may more generally be asymmetric. Considering symmetric interaction in the dynamics is only an approximation that may
simplify theoretical analysis, but which may fail to capture important phenomena occurring in real systems. 
For example, asymmetric interaction leads to novel features such as 
families of travelling wave states~\cite{Iatsenko:2013,Petkoski:2013}, glassy states
and super-relaxation~\cite{Iatsenko:2014}, and so forth, and has been invoked to
discuss coupled circadian neurons~\cite{Gu:2016}, dynamic
interactions~\cite{Yang:2020,Sakaguchi:1988}, etc. A
generalization of the Kuramoto model~(\ref{eq:eom-0}) that accounts for
asymmetric interaction is the so-called Sakaguchi-Kuramoto model, with
the dynamics described by the equation of motion~\cite{Sakaguchi:1986} 
\begin{equation}
\frac{{\rm d}\theta_j}{{\rm d}t}=\omega_j+\frac{K}{N}\sum_{k=1}^N
\sin(\theta_k-\theta_j +\alpha),
\label{eq:eom-SK}
\end{equation}
where $0\le\alpha < \pi/2$ is the so-called phase lag parameter. It is
now easily seen that owing to the presence of an $\alpha \ne 0$, the influence on the $j$-th oscillator due to the
$k$-th one is not any more equal in magnitude to the influence on the $k$-th
oscillator due to the $j$-th one. As has been the
case with
the Kuramoto model, the model~(\ref{eq:eom-SK}) and its
variants have been successfully invoked to study a variety of dynamical
scenarios, including, to name a few, disordered Josephson series
array~\cite{Wiesenfeld:1996}, time-delayed
interactions~\cite{Yeung:1999}, hierarchical populations of coupled
oscillators~\cite{Pikovsky:2008}, chaotic
transients~\cite{Wolfrum:2011}, dynamics of pulse-coupled
oscillators~\cite{Pazo:2014}, etc.  We note in passing that asymmetric interaction arises in the Kuramoto model with spatially heterogeneous time-delays in the interaction, see Ref.  \cite{Skardal:2018}. Let us note that both the dynamics~(\ref{eq:eom-0})
and~(\ref{eq:eom-SK}) satisfy phase-shift symmetry, whereby rotating
every phase by an
arbitrary angle same for all leaves the dynamics invariant. In
particular, one may implement the transformation
$\theta_j(t) \to \theta_j(t)+\omega_0 t~\forall~j,t$, which is
tantamount to viewing the dynamics in a frame rotating uniformly with
frequency $\omega_0$ with respect to an inertial frame, e.g., the
laboratory frame. 

The paper is organized as follows.  In the following section, we describe the model of study, Eq. (\ref{eq:eom}), and summarize our results obtained later in the paper. In Section~\ref{sec:analysis}, we
present our Ott Antonsen ansatz-based analysis of the dynamics~(\ref{eq:eom}) and the existence and stability of all the different possible
states at long times. In Section~\ref{sec:numerics},
we discuss our analytical vis-\`{a}-vis numerical results. Finally, in
Section~\ref{sec:conclusions}, we draw
our conclusions.

\section{Model of study and Summary of results}

In this work, we study a generalization of the Sakaguchi-Kuramoto dynamics~(\ref{eq:eom-SK})
by including an interaction term that explicitly breaks the phase-shift
symmetry of the dynamics. To this end, we consider the
following set of $N$ coupled nonlinear differential equations:
\begin{equation}
\frac{{\rm d}\theta_j}{{\rm d}t}=\omega_j+\frac{1}{N}\bigg[\epsilon_1
\sum_{k=1}^N \sin(\theta_k-\theta_j+\alpha)+
\epsilon_2\sum_{k=1}^N
\sin(\theta_k+\theta_j+\alpha)\bigg],
\label{eq:eom}
\end{equation}
where the real parameters $\epsilon_{1,2}$ denote the coupling
constants. Equation (\ref{eq:eom}) may be obtained as a result of  phase reduction analysis of a model of all-to-all-coupled Stuart-Landau oscillators with a symmetry-breaking form of interaction ~\cite{cj23,cj3}.

Equation~(\ref{eq:eom}) may also be written as 
\begin{equation}
\frac{d\theta_i}{dt}=\omega_i+\sum_{k=1}^{2}Q_k(\theta_i) \frac{f_k}{N}\sum_{j=1}^{N}P_k(\theta_j),
\label{eq:wf}
\end{equation}
where we have $f_1=\epsilon_1+\epsilon_2$,  $f_2=\epsilon_2-\epsilon_1$, $P_1(\theta)=\cos(\theta)$, $P_2(\theta)=\sin(\theta)$, $Q_1(\theta)=\sin(\theta+\alpha)$ and $Q_2(\theta)=\cos(\theta+\alpha)$. Equation (\ref{eq:wf}) resembles the Winfree model \cite{winfree},  so we may consider our model of study as  a variant of a Winfree-type system.  The Winfree model played a seminal role in the field of collective synchrony, inspiring the Kuramoto model as well as promoting recent advances in theoretical neuroscience~\cite{wf-3,wf-4}. In spite of the simplifying assumptions of  the  Winfree  model,  namely, that of uniform  all-to-all  weak  coupling, analytical  solutions  have  been  found  only  recently  using the  Ott-Antonsen  ansatz~\cite{wf-5,Pazo:2014},  see  also~\cite{wf-6}.

Note that on putting $\epsilon_2=0$ in
Eq.~(\ref{eq:eom}), one recovers the dynamics of the Sakaguchi-Kuramoto
model~(\ref{eq:eom-SK}) on identifying $\epsilon_1$ with the parameter $K
\ge 0$ in the latter, and so we take $\epsilon_1$ to be positive. Here,
we consider $\epsilon_2$ to satisfy $\epsilon_2 \ge 0$. The
dynamics~(\ref{eq:eom}) has phase-shift symmetry only with the choice
$\epsilon_2=0$, and so the $\epsilon_2$-term acts a
phase-shift-symmetry-breaking interaction (With $\epsilon_2 \ne
0$, the only case of phase-shift symmetry is with respect to the transformation $\theta_j \to
\theta_j+\pi~\forall~j$.). In contrast to the
Sakaguchi-Kuramoto model, the dynamics~(\ref{eq:eom}) is not invariant
with respect to the transformation $\theta_j \to \theta_j + \omega_0
t~\forall~j,t$ owing to the presence of the $\epsilon_2$-term.
Consequently, the mean $\omega_0$ of the frequency distribution
$g(\omega)$ would have a significant influence on
the dynamics~(\ref{eq:eom}), and cannot be gotten rid of by viewing the
dynamics in a frame rotating uniformly
with frequency $\omega_0$ with respect to the laboratory frame, as is
possible with the Sakaguchi-Kuramoto model. We note in passing that the so-called active rotators may also be described by phase oscillators.  Active rotators are well-established paradigms for excitable systems,  and possess a term that breaks the phase-shift symmetry ~\cite{Sakaguchi:1988,AR2,AR3}.

We will consider in this
work a unimodal $g(\omega)$. Specifically, we will consider a Lorentzian
$g(\omega)$:
\begin{equation}
g(\omega)=\frac{\gamma}{\pi((\omega-\omega_0)^2+\gamma^2)};~~\gamma >0.
\label{eq:lor}
\end{equation}

Analysis of synchronization in the context of the Kuramoto class of
models involves introducing a complex-valued order parameter given
by~\cite{Kuramoto:1984}
\begin{equation}
        z(t)=r(t)e^{\im\psi(t)}\equiv \frac{1}{N}\sum_{j=1}^N
e^{\im\theta_j(t)}.
\label{eq:r}
\end{equation}
The quantity $r(t);~0 \le r(t) \le 1$, measures the amount
of synchrony present in the system at time $t$, while
$\psi(t) \in [-\pi,\pi]$ gives the average phase.
Considering the limit $N \to \infty$, both the dynamics~(\ref{eq:eom-0})
and (\ref{eq:eom-SK}) allow for a stationary state to exist at long
times. By stationary state is meant a state with time
independent $z$. In such a state, the system
may be found in a synchronized or an incoherent state. The two
states are characterized by the time-independent value that $r(t)$
assumes at long times, namely, a zero value in the incoherent state and
a non-zero value in the synchronized state.

Considering the limit $N \to \infty$, this work aims at a detailed
characterization of the long-time ($t\to \infty$) limit of the
dynamics~(\ref{eq:eom}) and understanding of what new features with
respect to the Sakaguchi-Kuramoto model are
brought in by the introduction of the phase-shift-symmetry-breaking
$\epsilon_2$-term. An asymmetric
phase-shift-symmetry-breaking interaction, such as the one being
considered by us, is an hitherto unexplored theme in the context of the
Sakaguchi-Kuramoto model, and raises many queries: Does the introduction of the
$\epsilon_2$-term suffice to allow for the dynamics to have a stationary
state, and if so, what is
the nature of the stationary state? What is the range of parameter values
for which one has a synchronized stationary state? Most importantly,
what is the complete bifurcation diagram in the
($\epsilon_1,\epsilon_2$)-plane?

Recently, the dynamics of the Kuramoto
model in presence of phase-shift-symmetry-breaking symmetric interaction
was shown to unveil a
rather rich bifurcation diagram with existence of both stationary and
non-stationary synchronized states~\cite{Chandru:2020}; the model
studied therein corresponds to the dynamics~(\ref{eq:eom}) with $\alpha$
set to zero. It is therefore pertinent that we summarize right at the
outset what new features does the dynamics~(\ref{eq:eom}) offer with
respect to (i) the Sakaguchi-Kuramoto model (i.e., the case
$\epsilon_2=0$ in Eq.~(\ref{eq:eom})), and (ii) the Kuramoto model with
an additional phase-symmetry-breaking symmetric interaction (i.e., the case
$\alpha=0$ in Eq.~(\ref{eq:eom})). We focus on unimodal frequency
distributions. As is well known, the Sakaguchi-Kuramoto model shows
either an incoherent ($r(t)$ vanishes as $t \to \infty$) or a
synchronized ($r(t)$ as $t\to \infty$ has a nonzero value, and moreover,
$r(t)$ does not oscillate as a function of time)
state~\cite{Sakaguchi:1986}. These states are
observed only in a co-rotating frame rotating uniformly with frequency
$\omega_0$ with respect to the laboratory frame. As one tunes the
strength of $\epsilon_1$, one has typically a continuous
synchronization transition between an incoherent and a synchronized state,
though for certain specific unimodal
frequency distributions and carefully chosen phase lag $\alpha$, one may
observe more than one non-trivial synchronization
transitions~\cite{Oleh-SK-1,Oleh-SK-2}. Now,
coming to the model (ii), it has been observed that in addition to the
incoherent and the synchronized state, the dynamics may also exhibit
what we called a standing wave state, namely, a state in which $r(t)$ at
long times oscillates as a function of time, yielding a non-zero
time-independent time average. Note that for the model (ii), all the
three mentioned states are observed in the laboratory frame itself.
Moreover, the bifurcation diagram in the
$(\epsilon_1,\epsilon_2)$-plane exhibits a continuous transition between
the incoherent and the standing wave state, but instead a first-order
transition between the incoherent and the synchronized state, and
between the standing wave and the synchronized state. The latter fact
implies regions of metastability or coexistence between the incoherent and the synchronized state, and
between the standing wave and the synchronized state. In the light of
the aforementioned facts, we now summarize the relevant features of the
bifurcation diagram in the $(\epsilon_1,\epsilon_2)$-plane for the
model~(\ref{eq:eom}) that we report in this work. The model exhibits in
the laboratory frame the incoherent (IC), the synchronized and the standing
wave state; for better highlighting of the differences between the last
two states, we call them the oscillation death (OD) state and the
oscillatory synchronized (OS) state,  respectively.  However, in contrast to the model
(ii), we now have multistability/coexistence between all the different
pair of states, that is, there are IC-OD, OS-OD, IC-OS coexistence
regions, and in addition, a very peculiar and striking three-state-coexistence
region between IC-OS-OD. We believe that this three-state coexistence is new to the
literature on Kuramoto and Sakaguchi-Kuramoto models. The highlight
of our work is thus the revelation that asymmetric interactions lead to a
very rich bifurcation diagram when compared to the original
Sakaguchi-Kuramoto model and the Kuramoto model with
an additional phase-symmetry-breaking symmetric interaction (i.e., the case
$\alpha=0$ in Eq.~(\ref{eq:eom})). In particular, with respect to models
(i) and (ii), all the possible transitions in
the model~(\ref{eq:eom}) are first-order, and, moreover, a new
three-state-coexistence region is observed; we may conclude therefore that
introducing a phase-shift-symmetry-breaking interaction in the
Sakaguchi-Kuramoto model drastically and non trivially modifies the bifurcation diagram with respect to the Sakaguchi-Kuramoto
model and with respect to the model studied in
Ref.~\cite{Chandru:2020}.

The rest of the paper is devoted to a derivation of the aforementioned
results. For Lorentzian $g(\omega)$, Eq.~(\ref{eq:lor}), we use exact
analytical results derived by applying the so-called Ott-Antonsen ansatz, 
combined with numerical integration of the dynamics~(\ref{eq:eom}) for
large $N$, to
support the key result of our work, the bifurcation diagram of
Fig.~\ref{fig:phase-diagram}. The celebrated Ott-Antonsen ansatz is a remarkable
method of analysis introduced to study coupled oscillator systems, which
allows to rewrite in the limit $N \to \infty$ the dynamics of coupled
networks of phase oscillators in terms of a few collective variables~\cite{Ott:2008,Ott:2009}. 

%%%%%%%%%%%%%%%%%%%%%%%%%%%%%%%%%%%%%%%%%%%%%%%%%%%%%%%%%%%%%%%%%%%%%%%%%%%%%%%%%%%%%%%%%%%%
\section{Analysis of the dynamics~(\ref{eq:eom}): The Ott-Antonsen ansatz}
\label{sec:analysis}

We now analyse the dynamics~(\ref{eq:eom}) in the limit
$N \to \infty$ by employing the Ott-Antonsen ansatz. In this limit, the dynamics
may be characterized by defining a single-oscillator
distribution function $f(\theta,\omega,t)$, which is such that $f(\theta,\omega,t)$
is the probability density of finding an oscillator with natural frequency
$\omega$ and phase $\theta$ at time $t$. The distribution is
$2\pi$-periodic in $\theta$ and is moreover normalized as
\begin{equation}
\int_0^{2\pi} {\rm
        d}\theta~f(\theta,\omega,t)=g(\omega)~\forall~\omega,t.
\label{eq:norm}
\end{equation}

The evolution of $f$ is given by the continuity equation describing the
conservation of number of oscillators of a given frequency under the
dynamics~(\ref{eq:eom}):
\begin{equation}
\frac{\partial f}{\partial t}+\frac{\partial }{\partial \theta}(fv)=0,
\label{eq:continuity-eqn}
\end{equation}
where $v(\theta,\omega,t) \equiv {\rm d}\theta/{\rm d}t$ is the angular
velocity at position $\theta$ at time
$t$. From Eq.~(\ref{eq:eom}), we get
\begin{align}
        v(\theta,\omega,t)=\omega+&\frac{\epsilon_1}{2\im}[ze^{-\im(\theta-\alpha)}-z^{\star}e^{\im(\theta-\alpha)}]\nonumber\\&-\frac{\epsilon_2}{2\im}[ze^{\im(\theta+\alpha)}-z^{\star}e^{-\im(\theta+\alpha)}],
\label{eq:v-theta}
\end{align}
where $z=z(t)$ is obtained as the $N \to \infty$-limit generalization of
Eq.~(\ref{eq:r}):  
\begin{equation}
        z= \int {\rm d}\omega \int {\rm d}\theta~f(\theta,\omega,t)e^{\im\theta}.
\label{eq:-z-Ntoinfty}
\end{equation}
In their seminal papers~\cite{Ott:2008,Ott:2009}, Ott and Antonsen pointed
out that all the attractors of the Kuramoto model and its many variants
and for the case of the Lorentzian $g(\omega)$, Eq.~(\ref{eq:lor}), may be found by using the ansatz
\begin{equation}
f(\theta,\omega,t)=\frac{g(\omega)}{2\pi}\left[1+\sum_{n=1}^\infty
\left(f_n(\omega,t) e^{\im n\theta}+{\rm c.c.}\right)\right].
\label{eq:f-Fourier}
\end{equation}
Here, c.c. stands for a term obtained by complex conjugation of the
first term within the brackets occurring in the sum, and 
\begin{equation}
f_n(\omega,t)=\left[a(\omega,t)\right]^n,
\label{eq:OA}
\end{equation}
with arbitrary $a(\omega,t)$ satisfying $|a(\omega, t)| < 1$, together with the requirements that
$a(\omega, t)$ may be analytically continued to the whole of the
complex-$\omega$ plane, it has no singularities in the lower-half
complex-$\omega$ plane, and $|a(\omega, t)| \to 0$ as ${\rm
Im}(\omega) \to -\infty$.

Using the ansatz~(\ref{eq:f-Fourier}) in Eq.~(\ref{eq:continuity-eqn}), we straightforwardly get
\begin{equation}
\frac{\partial a}{\partial t}+\im\omega a+\frac{\epsilon_1}{2}(za^2
e^{\im\alpha}-z^\star e^{-\im\alpha})+\frac{\epsilon_2}{2}(z
e^{\im\alpha}-z^\star a^2 e^{-\im\alpha})=0,
\label{eq:a-dynamics}
\end{equation}
where we have
\begin{equation}
z^\star=\int_{-\infty}^{\infty}a(t,\omega)g(\omega)d\omega=a((\omega_0-{\rm
i}\gamma),t),
\label{eq:zstar}
\end{equation}
obtained by using in Eq.~(\ref{eq:-z-Ntoinfty}) the
ansatz~(\ref{eq:f-Fourier}) and Eq.~(\ref{eq:lor}) for $g(\omega)$, then converting the integral therein into
a contour integral, and finally evaluating the contour integral. Equation~(\ref{eq:a-dynamics}) may then be rewritten as
\begin{align}
        \frac{\partial z}{\partial t}-{\rm
i}(\omega_0+\im&\gamma)z+\frac{\epsilon_1}{2}\big[|z|^2ze^{-\im\alpha}-ze^{\im\alpha}\big]\nonumber\\&+\frac{\epsilon_2}{2}\big[z^\star e^{-\im
\alpha}-z^3 e^{\im\alpha}\big]=0.
\label{eq:z-dynamics}
\end{align}

The above equation expressed in terms of the quantities $r$ and
$\psi$, Eq.  (\ref{eq:r}),  gives the following two coupled equations:
\begin{align}
        &\frac{{\rm d}r}{{\rm d}t}=-\gamma
        r-\frac{r}{2}(r^2-1)(\epsilon_1\cos(\alpha)-{\epsilon_2}\cos(2\psi+\alpha)),\nonumber\\
\label{eq:r-dynamics}\\
&\frac{{\rm d}\psi}{{\rm
d}t}=\omega_0+\frac{1}{2}(r^2+1)(\epsilon_1\sin(\alpha)+{\epsilon_2}\sin(2\psi+\alpha)).\nonumber
\end{align}
The above equations constitute the Ott-Antonsen-ansatz-reduced order parameter
dynamics corresponding to the dynamics~(\ref{eq:eom}) in the limit $N
\to \infty$ and for the case of the Lorentzian $g(\omega)$,
Eq.~(\ref{eq:lor}). Let us remark that these equations are invariant under the transformation
$(r,\psi) \to (r,\psi+\pi)$, which may be traced back to the fact that
the original dynamics~(\ref{eq:eom}) is invariant under the
transformation $\theta_j \to \theta_j+\pi~\forall~j$.

Note that for $\epsilon_2=0$ and $\alpha=0$, when one has the Kuramoto
model, the two equations in~(\ref{eq:r-dynamics}) are decoupled, and the equations then allow for only uniform
rotation of $\psi$ with frequency $\omega_0$. The case of our
interest, namely, $\epsilon_2 \ne 0$ and $\alpha \ne 0$, is analysed in
the subsection that follows. It would prove convenient for the
discussion presented therein to define the time average of $r(t)$ in the long-time ($t \to
\infty$) limit as
\begin{equation}
R\equiv \lim_{t \to \infty} \frac{1}{\tau}\int_t^{t+\tau} {\rm
d}t'~r(t').
\label{eq:R-definition}
\end{equation}

%%%%%%%%%%%%%%%%%%%%%%%%%%%%%%%%%%%%%%%%%%%%%%%%%%%%%%%%%%%%%%%%%%%%%%%%%%%%%%%%%%%%%%%%%%%%
%%%%%%%%%%%%%%%%%%%%%%%%%%%%%%%%%%%%%%%%%%%%%%%%%%%%%%%%%%%%%%%%%%%%%%%%%%%%%%%%%%%%%%%%%%%%
\subsection{Analysis of the Ott-Antonsen-ansatz-based dynamics}
\label{sec:OA-results}

Below we discuss the various states obtained in the long-time ($t \to
\infty$) limit of the dynamics~(\ref{eq:z-dynamics}), equivalently,
Eq.~(\ref{eq:r-dynamics}). 
%%%%%%%%%%%%%%%%%%%%%%%%%%%%%%%%%%%%%%%%%%%%%%%%%%%%%%%%%%%%%%%%%%%%%%%%%%%%%%%%%%%%%%%%%%%%
%%%%%%%%%%%%%%%%%%%%%%%%%%%%%%%%%%%%%%%%%%%%%%%%%%%%%%%%%%%%%%%%%%%%%%%%%%%%%%%%%%%%%%%%%%%%
%%%%%%%%%%%%%%%%%%%%%%%%%%%%%%%%%%%%%%%%%%%%%%%%%%%%%%%%%%%%%%%%%%%%%%%%%%%%%%%%%%%%%%%%%%%%
\subsubsection{Incoherent (IC) state}
\label{subsec:iss}

The incoherent (IC) state is characterized by time independent $z$ satisfying
$z=z^\star=0$ (thus representing a stationary state of the
dynamics~(\ref{eq:r-dynamics})); correspondingly, one has $r=0$, and hence,
$R=0$. The linear stability of such a state is determined by linearising
Eq.~(\ref{eq:z-dynamics}) around $z=0$, by writing $z$ as $z=u$ with
$|u|~ \ll$1. We obtain 
\begin{equation}
\frac{\partial u}{\partial t}+(\gamma-\im \omega_0)u-\frac{\epsilon_1}{2}(ue^{-\im \alpha})+\frac{\epsilon_2}{2}(u^{\star}e^{\im \alpha})=0.
\label{eq:u-dynamics}
\end{equation}
Writing $u=u_x + \im u_y$ yields
\begin{equation}
\frac{\partial }{\partial t}
\begin{bmatrix}
u_x \\
u_y
\end{bmatrix}
=M
\begin{bmatrix}
u_x \\
u_y
\end{bmatrix};\\
\end{equation}
\begin{equation}
~~M \equiv \begin{bmatrix}
-\gamma+\cos(\alpha)\big[\frac{\epsilon_1}{2}-\frac{\epsilon_2}{2}\big] & -\omega_0-\sin(\alpha)\big[\frac{\epsilon_1}{2}-\frac{\epsilon_2}{2}\big] \\\\
~\omega_0+\sin(\alpha)\big[\frac{\epsilon_1}{2}+\frac{\epsilon_2}{2}\big]& -\gamma+\cos(\alpha)\big[\frac{\epsilon_1}{2}+\frac{\epsilon_2}{2}\big] \\
\end{bmatrix}.
\label{eq:M-matrix} \nonumber
\end{equation}
The matrix $M$ has eigenvalues
\begin{equation}
\lambda_{1,2}=\frac{-2\gamma+\epsilon_1 \cos(\alpha)\pm \sqrt{\Delta}}{2},
\label{eq:M-eigenvalues}
\end{equation}
with
$\Delta=(\epsilon_2^2-\epsilon_1^2 \sin^2\alpha-4\omega_0^2-4\epsilon_1
\omega_0 \sin(\alpha))$. Note that we have $\lambda_1 > \lambda_2$. The
stability threshold for the IC is then obtained by analysing $\lambda_1$
as a function of $\epsilon_1$ and $\epsilon_2$, and asking for a given
$\epsilon_2$ the particular value of $\epsilon_1$ at which $\lambda_1$
vanishes. One thus obtains the stability threshold as
 \begin{align}
&\epsilon_1=2 \gamma  \sec (\alpha),~~~~~~~~~~~~~~~~   ~~~~~~~~~~~~ \;\;\;
         \mbox{for}\;\; \Delta \le 0, \nonumber\\
&\epsilon_1=2 \gamma  \cos (\alpha )-2 \omega_0  \sin (\alpha
)-P
\,,~~~~~ \;\;\; \mbox{for}
\;\; \Delta>0.\nonumber \\
\label{eq:stability-ISS}
\end{align}
Here, we have
\begin{align}
P=\sqrt{\epsilon_2^2-4 \gamma ^2 \sin^2 (\alpha )-4
	\omega_0^2 \cos^2 (\alpha )-4 \gamma  \omega  \sin (2 \alpha)}.
	\end{align}
%%%%%%%%%%%%%%%%%%%%%%%%%%%%%%%%%%%%%%%%%%%%%%%%%%%%%%%%%%%%%%%%%%%%%%%%%%%%%%%%%%%%%%%%%%%%
%%%%%%%%%%%%%%%%%%%%%%%%%%%%%%%%%%%%%%%%%%%%%%%%%%%%%%%%%%%%%%%%%%%%%%%%%%%%%%%%%%%%%%%%%%%%
%%%%%%%%%%%%%%%%%%%%%%%%%%%%%%%%%%%%%%%%%%%%%%%%%%%%%%%%%%%%%%%%%%%%%%%%%%%%%%%%%%%%%%%%%%%%
\subsubsection{Oscillatory Synchronized (OS) state}
\label{subsec:dss}

The oscillatory synchronized (OS) state is characterized by $z$ that is
time dependent (thus characterizing a non-stationary state of the
dynamics~(\ref{eq:r-dynamics})). In this state, the order
parameter $r(t)$ oscillates as a function of $t$, but
nevertheless yields a non-zero time-independent time average, $R \ne 0$.
It is thus distinct from the non-oscillatory synchronized state (i.e., an
oscillation death (OD) state, see below) that is also possible as a
long-time solution of the dynamics~(\ref{eq:r-dynamics}). For the OS,
one has $z$ that is time independent (thus characterizing a stationary state of the
dynamics~(\ref{eq:r-dynamics})) and correspondingly, both $r(t)$
and $R$ have non-zero time-independent values, but the former does not oscillate as a function of time. 
Deriving stability conditions for the OS does not prove easy, unlike
the IC. Hence, we analysed using Eq.~(\ref{eq:r-dynamics})
the OS stability using the numerical package XPPAUT~\cite{xpp}. The
analysis is pursued by expressing Eq.~(\ref{eq:r-dynamics}) in the Cartesian coordinates $x=r\cos \psi,~y=r\sin \psi;~~x,y \in [-\infty,\infty]$.
Our analysis reveals that the OS represents a stable limit cycle in
the $(x,y)$-plane. In Section~\ref{sec:numerics}, we will present XPPAUT-generated phase-space portraits in the $(x,y)$-plane for not only the IC, the
OS and the OD state, but also for regions in parameter space allowing
for coexistence of two or more states. Let us remark in passing that
owing to the invariance of the dynamics~(\ref{eq:r-dynamics}) under the transformation
$(r,\psi+\pi)$, it follows that the phase portrait would be
symmetric under $(x,y) \leftrightarrow (-x,-y)$.

%%%%%%%%%%%%%%%%%%%%%%%%%%%%%%%%%%%%%%%%%%%%%%%%%%%%%%%%%%%%%%%%%%%%%%%%%%%%%%%%%%%%%%%%%%%%
%%%%%%%%%%%%%%%%%%%%%%%%%%%%%%%%%%%%%%%%%%%%%%%%%%%%%%%%%%%%%%%%%%%%%%%%%%%%%%%%%%%%%%%%%%%%
%%%%%%%%%%%%%%%%%%%%%%%%%%%%%%%%%%%%%%%%%%%%%%%%%%%%%%%%%%%%%%%%%%%%%%%%%%%%%%%%%%%%%%%%%%%%
\subsubsection{Oscillation Death (OD) state}
\label{subsec:sss}

Considering the dynamics~(\ref{eq:r-dynamics}), we now explore the
possibility of existence of the oscillation death (OD) state.
Requiring in the dynamics that $r$ and $\psi$ have time-independent non-zero
values so that the left hand side of the two equations
 may be set to zero, we obtain for
the OD the two coupled equations
\begin{align}
&\frac{\epsilon_2}{2}\cos(2\psi+\alpha)(r^2-1)=\gamma +\frac{\epsilon_1}{2}\cos(\alpha)(r^2-1),\nonumber\\
&\frac{\epsilon_2}{2}\sin(2\psi+\alpha)(r^2+1)=-\omega
_0-\frac{\epsilon_1}{2}\sin(\alpha)(r^2+1). \nonumber \\
\label{eq:SSS-dynamics}
\end{align}
The above equations give the following solutions for stationary $r$ and
$\psi$:
\begin{align}
&\frac{\epsilon_2^2}{4}=\bigg(\frac{\gamma}{r^2-1}+\frac{\epsilon_1}{2}\cos(\alpha)\bigg)^2+~\bigg(\frac{\omega
_0}{r^2+1}+\frac{\epsilon_1}{2}\sin(\alpha)\bigg)^2,\nonumber\\
        &\tan(2\psi+\alpha)=\frac{(r^2-1)(-\omega
        _0-\frac{\epsilon_1}{2}\sin(\alpha)(r^2+1))}{(r^2+1)(\gamma
        +\frac{\epsilon_1}{2}\cos(\alpha)(r^2-1))}.\nonumber
\\
\label{eq:stability-SSS}
\end{align}
We note in passing that in the context of our model,  OD refers to a state in which $r(t)$ at long times has a non-zero time-independent value,  while OD refers to quenching of oscillations in coupled dynamical networks \cite{DV}.

In the next section, we view the above results in the light of those obtained from numerical integration of the
dynamics~(\ref{eq:eom}). 

%%%%%%%%%%%%%%%%%%%%%%%%%%%%%%%%%%%%%%%%%%%%%%%%%%%%%%%%%%%%%%%%%%%%%%%%%%%%%%%%%%%%%%%%%%%%
\section{Numerical results}
\label{sec:numerics}

\begin{figure}[!ht]
	\hspace*{-1cm}
\includegraphics[width=9cm]{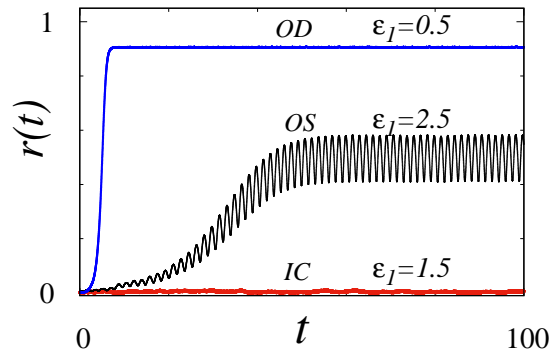}
\caption{Temporal behavior of $r(t)$ corresponding to the IC, the OS and the OD state obtained
        with the dynamics~(\ref{eq:eom}) for $\alpha=1.3$,
        $\epsilon_2=2.0$ and with varying values of $\epsilon_1$
        mentioned in the figure. The frequency distribution is the
        Lorentzian~(\ref{eq:lor}) with $\gamma=0.3$ and $\omega_0=0.6$.
        The data are obtained from numerical integration of the
        dynamics~(\ref{eq:eom}) with $N=10^5$.}
	\label{fig:1}
\end{figure}

\begin{figure}[!ht]
	\hspace*{-1cm}
\includegraphics[width=9cm]{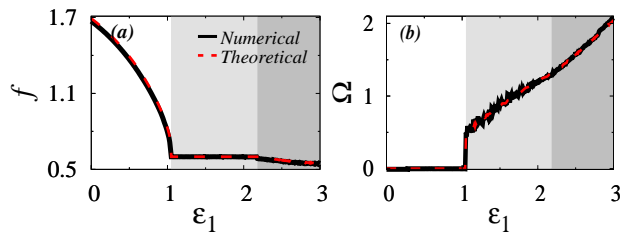}
\caption{The figure shows corresponding to Fig.  \ref{fig:1} the variation of the mean-ensemble frequency $f$ and the mean-field frequency $\Omega$ as a function of $\epsilon_1$.  Here,  we have compared numerical integration results with those based on the OA ansatz.  The different regimes are differentiated with different grey scales: IC and OS
represented by light and dark grey shades, respectively,  while OD is represented by
white shade.}
	\label{fig:1a}
\end{figure}

\begin{figure*}[!ht]
	\hspace*{-0.5cm}
	\includegraphics[width=12cm]{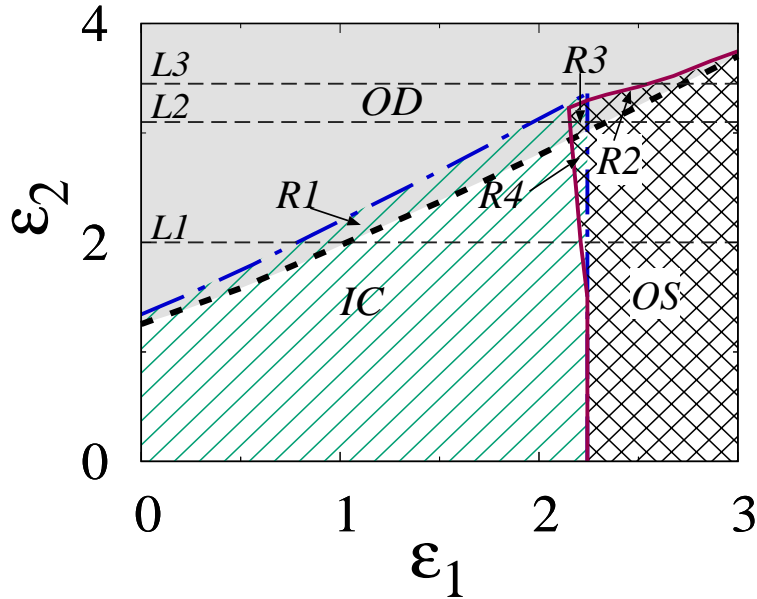}
	\caption{For the model~(\ref{eq:eom}) and considering for $\omega_j$'s the Lorentzian distribution~(\ref{eq:lor}) with $\gamma=0.3$ and $\omega_0=0.6$ , the figure depicts the
                two-parameter bifurcation diagram in the
		($\epsilon_1,\epsilon_2$)-plane for $\alpha=1.3$. We
                show here the various stable states and their boundaries, with bifurcation behavior
		observed as one crosses the
		different boundaries. The regions representing stable existence of the
		OD (Oscillation Death state), the IC (Incoherent 
		state) and the OS (Oscillatory Synchronized state) have been constructed
		by analysing the long-time numerical solution of the
		dynamics~(\ref{eq:eom}) for $N=10^4$, namely, by asking
                at given values of $\epsilon_1$ and
                $\epsilon_2$, the quantity $r(t)$ at long times represents which one of the
                three possible behavior depicted in Fig.~\ref{fig:1}. The blue
                dot-dashed line, the black dashed line, and the maroon
                continuous 
               line are stability boundaries of the IC, the OD and the
                OS, respectively, and have been obtained by using the
Ott-Antonsen-ansatz-based dynamics~(\ref{eq:r-dynamics}) and following
                the procedure of its analysis detailed in the text. The regions
                R1, R2, R3, and R4 represent multistability or
                coexistence between
                IC-OD, between OS-OD, between IC-OS-OD and between
                IC-OS, respectively.  The dashed
                lines L1, L2 and L3 indicate the values of $\epsilon_2$
                at which the plots of
                Figs.~\ref{fig:ts1} and~\ref{fig:at2}, reported later in
                the paper, are obtained.
            }
             
	\label{fig:phase-diagram}
\end{figure*}

\begin{figure*}[!ht]
    %  \hspace*{-1cm}
    \centering
	\includegraphics[width=13.5cm]{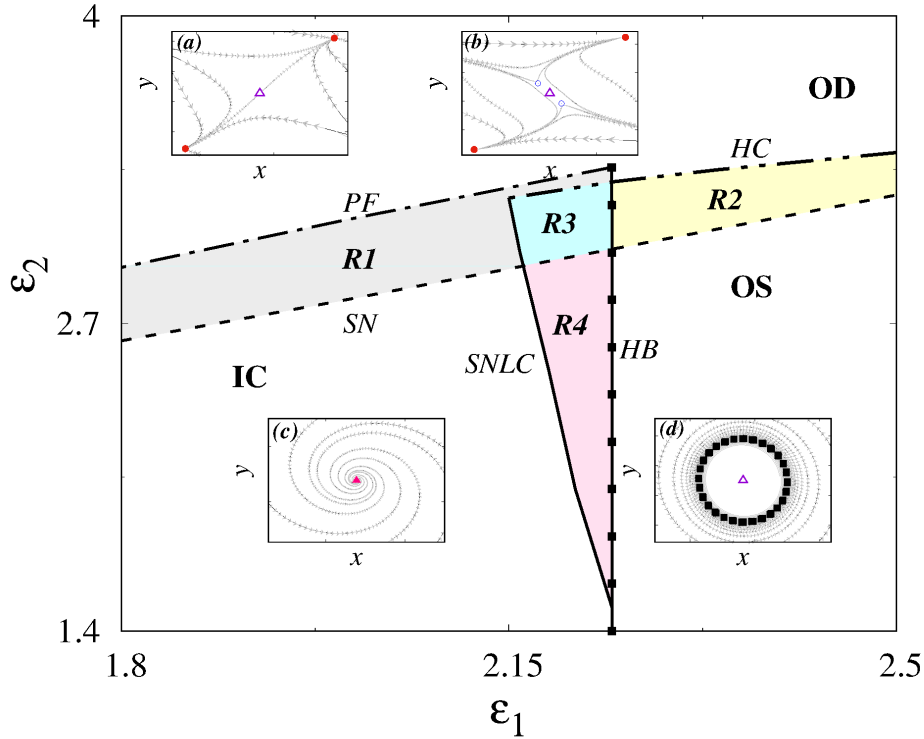}
        \caption{A zoom in on the bifurcation diagram in
        Fig.~\ref{fig:phase-diagram}. In the figure, we show that
the transitions between the IC, the OS, and the OD state are mediated by either a pitchfork
(PF, dot-dashed line),
or a saddle-node (SN, dotted line), or a saddle-node
        limit-cycle (SNLC, continuous line), or a Hopf (HB, continuous line containing filled
squares) or a homoclinic (HC, double-dot dashed line) bifurcation. Corresponding to the
        Ott-Antonsen-ansatz-based dynamics~(\ref{eq:r-dynamics}), the insets
        show the XPPAUT-generated phase-space portraits in the
        $(x,y)$-plane in the region of the bifurcation diagram in which
        they have been placed (red filled triangle: stable spiral, red filled circle:
        stable node, black
        filled square: stable limit cycle, purple open triangle:
        unstable fixed point, blue open circle: saddle). The nature of
bifurcations as mentioned above has also been obtained from the XPPAUT analysis of the
Ott-Antonsen-ansatz-based dynamics~(\ref{eq:r-dynamics})}
	\label{fig:b}
\end{figure*}

\begin{figure*}[!ht]
	\includegraphics[width=15cm]{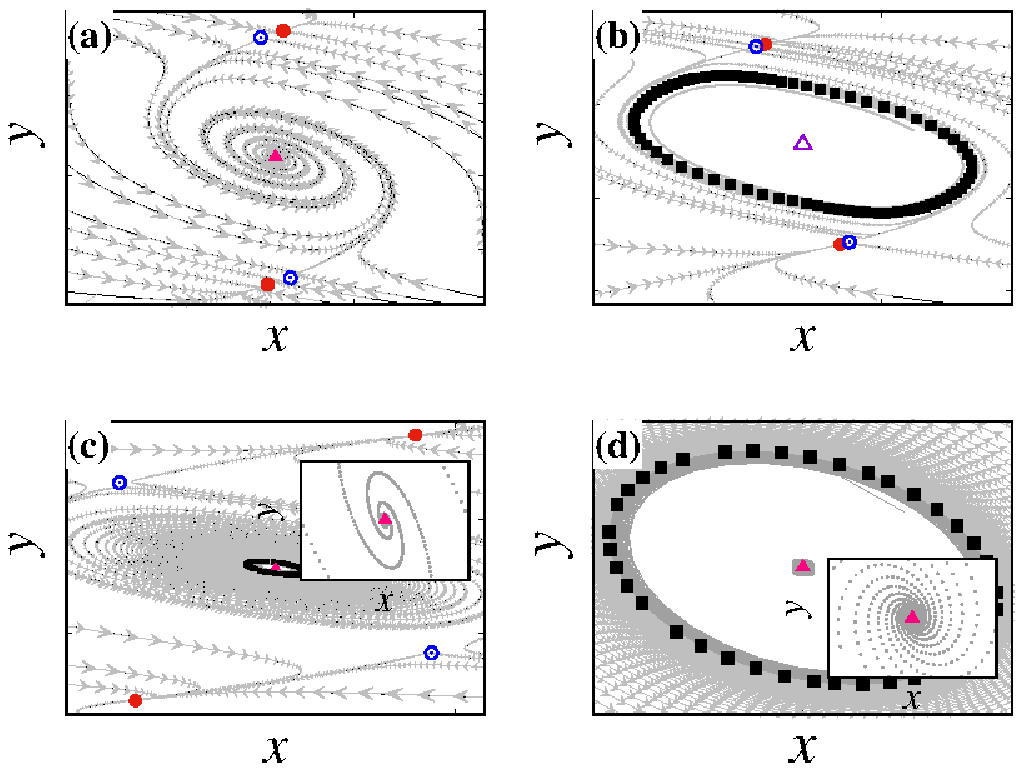}
	\caption{Corresponding to the
        Ott-Antonsen-ansatz-based dynamics~(\ref{eq:r-dynamics}), the figures
        show the XPPAUT-generated phase-space portraits in the
        $(x,y)$-plane for the different coexistence or multistability
        regions in the bifurcation diagram~\ref{fig:phase-diagram},
        namely, panel (a): R1; panel (b): R2; panel (c): R3; panel (d):
        R4. Legends: (i) red filled triangle: stable spiral, (ii) red filled circle:
        stable node, (iii) purple open triangle: unstable fixed point,
        (iv) black
        filled square: stable limit cycle, (v) blue open circle: saddle.}
	\label{fig:mb}
\end{figure*}

In the preceding section, we discussed the existence of the IC, the OS and
the OD state. Throughout this work, while discussing numerical integration
results for the dynamics~(\ref{eq:eom}), unless stated otherwise, we consider the number of
oscillators to be $N=10^5$ and the
natural frequency distribution to be given by the
Lorentzian~(\ref{eq:lor}) with $\gamma=0.3$ and $\omega_0=0.6$.
Numerical integration is
performed by employing a standard fourth-order Runge-Kutta integration
scheme with integration step size ${\rm d}t=0.01$.

In Fig. \ref{fig:1}, we have
plotted the temporal behavior of the IC, the OS and
the OD, for $\alpha=1.3$. The data are obtained from numerical integration of the dynamics~(\ref{eq:eom}). At fixed $\epsilon_2=2.0$, one has the
OD at $\epsilon_1=0.5$, the IC at $\epsilon_1=1.5$, and the OS at
$\epsilon_1=2.5$. As expected, we see that the three states are distinguished by the different
long-time temporal behavior of $r(t)$. Namely, for the IC, $r(t)$ takes the value zero in the $t \to \infty$ limit. 
In the OS, $r(t)$ as
$t \to \infty$ oscillates
in time. For the OD, $r(t)$ as $t \to \infty$ has a non-zero
time-independent constant value.
Figure \ref{fig:1a} shows as a function of $\epsilon_1$ the mean-ensemble frequency $f$ and the mean-field frequency $\Omega$ at long times.  For details of computation of these quantities, we refer the reader to Ref.  \cite{Chandru:2020}.

In order to demonstrate the influence of
$\alpha$ on the stability of the different states, we now discuss the
bifurcation diagram in the ($\epsilon_1,\epsilon_2$)-plane. Figure~\ref{fig:phase-diagram} shows the stable states in the
$(\epsilon_1,\epsilon_2)$-plane for the model~(\ref{eq:eom}) with
$\alpha=1.3$. Here the
states represent either the IC or the OS or the OD state. The regions representing stable existence of the IC, the OS and the OD state have been constructed by analysing the long-time numerical solution of the dynamics~(\ref{eq:eom}) for $N=10^4$, namely, by asking at given values of $\epsilon_1$ and $\epsilon_2$ the quantity $r(t)$ at long times represents which one of the three possible behavior depicted in Fig.~\ref{fig:1}.  
The stability boundary for the IC (blue dot-dashed 
line) is given by the Ott-Antonsen-ansatz-based
equation, Eq.~(\ref{eq:stability-ISS}).  The stability boundary for the OD (black dashed line)
is obtained from the Ott-Antonsen-ansatz-based
equation, Eq.~(\ref{eq:stability-SSS}). The procedure is as follows: For
given $\gamma$, $\omega_0$ and $\alpha$, we vary at fixed $\epsilon_2$ the value of
$\epsilon_1$ from high to low, and use the first equation in~(\ref{eq:stability-SSS}) to
ascertain the particular value of $\epsilon_1$ when for the first time
the equation admits a solution for $r$ in the range $0 < r \le 1$; this
particular value of $\epsilon_1$ gives the stability threshold of the OD
at the chosen value of $\epsilon_2$. The process is repeated for
different values of $\epsilon_2$.  The stability boundary for the OS (maroon continuous line) is obtained from the XPPAUT
analysis discussed in Section~\ref{subsec:dss} for the Ott-Antonsen-ansatz-based
                dynamics, Eq.~(\ref{eq:r-dynamics}). The detailed procedure
involves the following: For
given $\gamma$, $\omega_0$ and $\alpha$, we vary at fixed $\epsilon_2$ the value of
$\epsilon_1$ from low to high, and use the numerical solution of the 
Eq.~(\ref{eq:r-dynamics}) (rewritten in terms of $x=x(t)$ and $y=y(t)$) to
ascertain the particular value of $\epsilon_1$ when we first encounter
at long times a stable limit cycle in the $(x,y)$-plane. We then repeat
the process for different values of $\epsilon_2$.

\begin{figure}[!ht]
	\hspace*{-0.6cm}
\includegraphics[width=9cm]{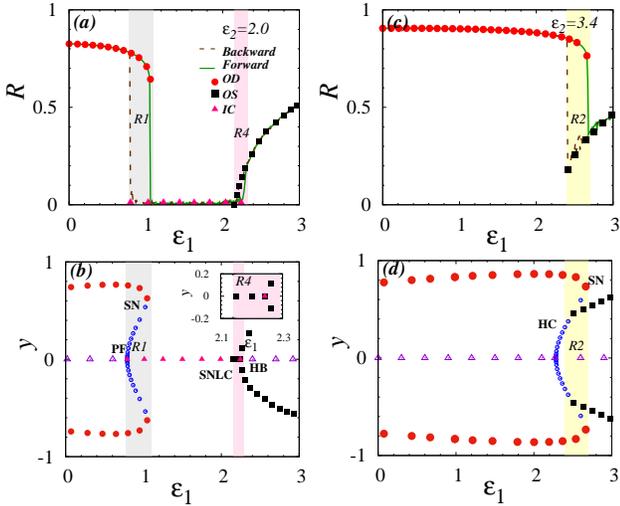}
\caption{The plots correspond to model~(\ref{eq:eom}) with $\alpha=1.3$,
and by considering for $\omega_j$'s the Lorentzian
distribution~(\ref{eq:lor}) with $\gamma=0.3$ and $\omega_0=0.6$. Panels
(a), (b) and (c), (d)  correspond to values of $\epsilon_2$
represented in Fig.~\ref{fig:phase-diagram} by dashed lines L1
($\epsilon_2=2.0$) and L3 ($\epsilon_2=3.4$), respectively. Panels (a) and (c) show as a function of
adiabatically-tuned $\epsilon_1$ the quantity $R$, namely, the
time-averaged order parameter in the long-time limit. The lines
correspond to results based on numerical integration of the
dynamics~(\ref{eq:eom}), while symbols correspond to the analysis of the
Ott-Antonsen-ansatz-based dynamics, Eq.~(\ref{eq:r-dynamics}), and are obtained as
detailed in the main text. The
symbols in panels (b) and (d) represent XPPAUT-generated bifurcation
plots for $y$ is a function of $\epsilon_1$ with fixed $\epsilon_2$ and based on the
Ott-Antonsen-ansatz-dynamics~(\ref{eq:r-dynamics}). All the panels show the
existence of the different states as well as multistability regions R1,
R2, and R4 (\textit{cf.} bifurcation diagram~\ref{fig:phase-diagram}).}
\label{fig:ts1}
\end{figure}

\begin{figure}[!ht]
	\hspace*{-1cm}
\includegraphics[width=8.5cm,height=11cm]{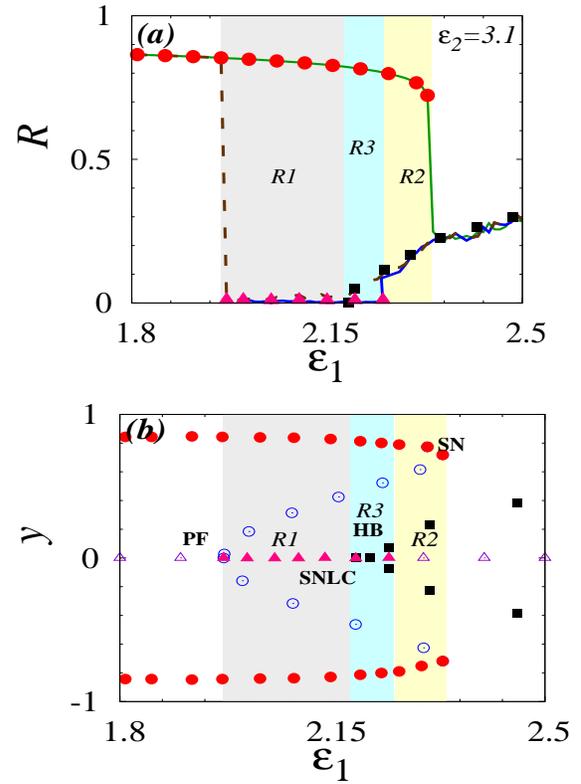}
\caption{The plots correspond to model~(\ref{eq:eom}) with
        $\alpha=1.3$, and by considering for $\omega_j$'s the Lorentzian distribution~(\ref{eq:lor}) with
		$\gamma=0.3$ and $\omega_0=0.6$.  
	 The panels correspond to the value $\epsilon_2=3.1$,
        represented as the straight line L2 in Fig.~\ref{fig:phase-diagram}.
		 Panel (a) shows as a function of adiabatically-tuned $\epsilon_1$ the
        quantity $R$, namely, the
        time-averaged order parameter in the long-time limit. The symbols in
panel (b) represent XPPAUT-generated bifurcation plots for $y$
as a function of $\epsilon_1$ with fixed $\epsilon_2$ and based on the
Ott-Antonsen-ansatz-dynamics~(\ref{eq:r-dynamics}). Here, panel (a) (respectively, panel (b))
has been obtained by following the same procedure as the one followed in
obtaining panels (a) and (c)
(respectively, panels (b) and (d)) of Fig.~\ref{fig:ts1}. The panels show the different states as well as the
multistability regions of R1, R2, R3. (\textit{cf.} bifurcation
diagram~\ref{fig:phase-diagram}).} 
	\label{fig:at2}
\end{figure}

Figure~\ref{fig:phase-diagram} shows in particular the presence of
multistability or coexistence regions R1, R2, R3 and R4; these represent 
multistability between IC-OD, between OS-OD,
between IC-OS-OD (coexistence of all the states) and between IC-OS, respectively. At a fixed $\epsilon_1$ and on tuning $\epsilon_2$ (or
vice versa), one observes bifurcations as one crosses the different boundaries. When compared with the bifurcation diagram in case of symmetric
interactions that either preserve or break phase-shift symmetry~\cite{Chandru:2020}, we see that presence of
asymmetry in the interaction leads to a very rich bifurcation diagram. { The dashed lines L1, L2, and L3 in
Fig.~\ref{fig:phase-diagram} indicate the values of $\epsilon_2$ for
which the plots in Figs.~\ref{fig:ts1} and ~\ref{fig:at2}, reported
later in the paper, are obtained.}
 Figure~\ref{fig:phase-diagram}, the bifurcation diagram of model~(\ref{eq:eom}), is the key result of our
work. 

Figure \ref{fig:b} shows further details of the bifurcation diagram, and in
particular, corresponding to the
Ott-Antonsen-ansatz-based dynamics~(\ref{eq:r-dynamics}), the XPPAUT-generated phase portraits in the $(x,y)$-plane for the
IC, the OS and the OD state. The OD corresponds to two symmetrically-placed
stable nodes at
nonzero $x$ and $y$ (red filled circle); the phase portrait, panel (a),
shows an unstable
fixed point at $x=0,~y=0$ (purple open triangle), while that in panel
(b) shows in addition to the unstable fixed point also two saddles (blue open circle). The IC represents a fixed point at
$x=0,~y=0$ that is a stable spiral (red filled triangle), so that trajectories in course of
evolution spiral in to
the fixed point, see panel (c). The OS corresponds to a stable limit cycle (black
filled square), so that
trajectories emanating from the unstable fixed point at $x=0,~y=0$ (purple open triangle) as well as
those from the region outside the limit cycle eventually approach the
limit cycle in course of evolution (see panel (d)). In the figure, we show that
the transitions between these states are mediated by either a pitchfork
(PF, dot-dashed line),
or a saddle-node (SN, dotted line), or a saddle-node limit-cycle (SNLC, continuous line), or a Hopf (HB, continuous line containing filled
squares), or a homoclinic (HC, double-dot dashed line) bifurcation. The nature of
bifurcations has been obtained from the XPPAUT analysis of the
Ott-Antonsen-ansatz-based dynamics~(\ref{eq:r-dynamics}).

In Fig.~\ref{fig:mb}, we show the XPPAUT-generated phase-space
portraits in each of the coexistence regions R1, R2, R3, R4, based on the
Ott-Antonsen-ansatz-based dynamics~(\ref{eq:r-dynamics}). In (a),
corresponding to R1, we see the coexistence of IC (a stable spiral (red filled
triangle)) and the OD (two stable nodes (red filled circle)), together
with two
saddles (blue open circle).
In (b), corresponding to R2, the OS (a stable limit cycle (black filled
square)) coexists with the OD (two stable nodes (red filled circle)),
together with an unstable fixed point (purple open triangle) and two saddles (blue open circle). 
In (c), corresponding to R3, the OS (a stable limit cycle (black filled
square)) coexists with the IC (a stable spiral (red filled triangle))
and the OD (two stable nodes (red filled circle)), and additionally,
there are two saddles (blue open circle).
In (d), corresponding to R4, the OS (a stable limit cycle (black filled
square)) coexists with the IC (a stable spiral (red filled triangle))
(in this case, there is an unstable limit cycle (not shown) separating
the two behavior).

Now that we have seen the complete bifurcation diagram and the phase portraits
in the three states, the IC, the OS and the OD, and in regions of their
coexistence, we turn to a discussion of the behavior of the quantity
$R$ (the time-averaged value of $r(t)$ computed at long times, see
Eq.~(\ref{eq:R-definition})) as a function of adiabatically-tuned $\epsilon_1$ for representative
values of $\epsilon_2$, with the aim to see signatures of bifurcation. We first
let the system settle to the stationary state at a fixed value
of $\epsilon_1$, and then tune $\epsilon_1$ adiabatically in time from low
to high values while recording the value of $R$ in time; this
corresponds to forward variation of $\epsilon_1$. Subsequently, we tune
$\epsilon_1$ adiabatically in time from high to low values (backward
variation of $\epsilon_1$). Adiabatic tuning ensures that the system
remains in the stationary state at all times as the value of
$\epsilon_1$ changes in time. In Fig.~\ref{fig:ts1}, we consider two
values of $\epsilon_2$: Panels
(a) and (b) are for $\epsilon_2=2.0$, while panels (c) and (d) are for
$\epsilon_2=3.4$. Here, the lines
correspond to results based on numerical integration of the
dynamics~(\ref{eq:eom}), while symbols correspond to the analysis of the
Ott-Antonsen-ansatz-based dynamics, Eq.~(\ref{eq:r-dynamics}), and are obtained as follows:
For the IC, the Ott-Antonsen analysis predicts that the IC
state with $R=0$ exists only for $\epsilon_1$ larger than its threshold
                value given by Eq.~(\ref{eq:stability-ISS}).  For the OS, the quantity $R$ is obtained from the XPPAUT analysis of Eq.~(\ref{eq:r-dynamics}).  For the OD, the values of $R$ are obtained from numerical
        integration of Eq.~(\ref{eq:stability-SSS}).

In panel (a), we observe hysteresis, implying occurrence of the
coexistence region R1 between the OD and the IC and the coexistence
region R4 between the IC and the OS. This is indeed consistent with
Fig.~\ref{fig:phase-diagram} which shows that
$\epsilon_2=2.0$ allows for the existence of R1 and R4 regions. The
hysteresis seen in panel (c) implies occurrence of the
coexistence region R2 between the OD and the OS, and is again consistent with
Fig.~\ref{fig:phase-diagram} in which we see that
$\epsilon_2=3.4$ allows for the existence of R2 region. The symbols in
panels (b) and (d) represent XPPAUT-generated bifurcation plots for $y$
as a function of $\epsilon_1$ with fixed $\epsilon_2$ and based on the Ott-Antonsen-ansatz-dynamics~(\ref{eq:r-dynamics}), and are a
further confirmation of multistability and bifurcation behavior. 

In panel (b), we see that when the IC becomes unstable (i) with the decrease
of $\epsilon_1$, one observes a subcritical pitchfork bifurcation (PF)
to give rise to the OD: at the bifurcation point, two symmetric unstable branches (blue open circle, represents saddles) and one stable
branch corresponding to the IC (red filled triangle) go over to one
unstable branch (purple open triangle, represents unstable fixed points); (ii) with the increase of
$\epsilon_1$, one observes  a saddle-node limit
cycle bifurcation (SNLC) to give rise to a stable and an unstable limit
cycle coexisting with the IC; the IC loses its stability with increase
of $\epsilon_1$ via Hopf bifurcation (HB), which gives rise to the
stable OS. At HB, the stable and the unstable limit cycle born at SNLC bifurcation
collide with each other and disappear.
One stable branch
corresponding to the IC (red filled triangle) goes over at the
bifurcation point to a stable branch
                corresponding to the OS (black filled square: upper and lower
                branches in the figure correspond to the maximum and the minimum of
                the corresponding stable limit cycle). Next, when the OD becomes unstable with the
increase of $\epsilon_1$, one observes a saddle-node bifurcation (SN) to
give rise to the IC: at the bifurcation point, one
                stable branch corresponding to the OD (red filled circle) and one
                unstable branch (blue open circle, represent saddles) annihilate each other. 
On the other hand, when the OS becomes unstable with the
decrease of $\epsilon_1$, one observes a saddle-node limit-cycle
bifurcation (SNLC) to give rise to the IC: at the bifurcation point, a stable branch corresponding to the OS (black
                filled square: upper and lower
                branches in the figure correspond to the maximum and the minimum of
                the corresponding stable limit cycle) and an unstable
                branch (purple open triangle, represents unstable
                spirals) annihilate each other.

 In panel (d), we see that when the OD becomes
                        unstable with the increase
                of $\epsilon_1$, one observes a saddle-node bifurcation
                to give rise to the OS: at the bifurcation point, a stable branch corresponding to the OD
                (red filled circle) and an unstable branch (blue open
                circle, represent saddles)
                annihilate each other.
                Similarly, when the OS becomes
                        unstable with the decrease
                of $\epsilon_1$, one observes a homoclinic bifurcation:
                at the bifurcation point, 
                a stable branch corresponding to the OS (a stable limit
                cycle) and represented by black filled squares (upper and lower
                branch correspond respectively to the maximum
                and the minimum of the limit cycle) collides with an
                unstable branch (blue open circle, represents saddles).

In order to witness the R3 region, one has to choose the
value of $\epsilon_2$ carefully as R3 represents a very narrow region, see
Fig.~\ref{fig:phase-diagram}; we consider $\epsilon_2=3.1$ to witness the R3
region, see Fig.~\ref{fig:at2}. Here, panel (a) (respectively, panel (b))
has been obtained by following the same procedure as the one followed in
obtaining panels (a) and (c)
(respectively, panels (b) and (d)) of Fig.~\ref{fig:ts1}. Here, we see the regions R1 and R2 as
well. In panel (a), we see during the forward (adiabatic) variation of
$\epsilon_1$ and starting with the OD (red filled circle) that the state disappears
with the emergence of the OS (black filled square), while during the
backward variation, the OS
destabilizes to give birth to the IC (red filled triangle) (this
confirms the existence of the multistability region R2), and eventually
back to the OD. Choosing the IC
as the initial state shows during the
forward variation of $\epsilon_1$ a destabilization to give rise to the OS and during the backward variation a destabilization to give rise to the IC and eventually to the
OD (this last behavior confirms the existence of the region R1). This establishes R3 as the region in
which the OD, the IC and the OS coexist. In panel (b), we see that when the OD becomes unstable with the
increase of $\epsilon_1$, one observes a saddle-node bifurcation (SN) to
the OS: at the bifurcation point, one stable branch corresponding to the OD (red filled circle) and one unstable branch (blue open circle, represent saddles) annihilate each other. When the OS becomes unstable with the decrease of $\epsilon_1$, it bifurcates (i) first as a Hopf bifurcation (HB) to the IC, and (ii) then as a saddle-node limit-cycle bifurcation to the IC. The IC so obtained undergoes with
                further decrease of $\epsilon_1$ a pitchfork bifurcation (PF) to the OD.

 \begin{figure}[!ht]
	\hspace*{-0.60cm}
 \includegraphics[width=9cm,height=8.2cm]{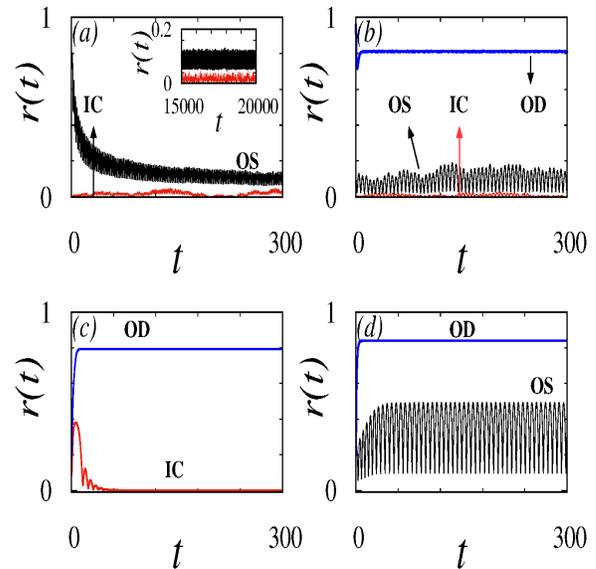}
 \caption{Temporal behaviour of $r(t)$ showing expected
 long-time stable coexistence of the different states under the
 dynamics~(\ref{eq:eom}) initiated with the respective states. The
 parameters $\omega_0$, $\gamma$ and $\alpha$ have the same values as in
 Fig.~\ref{fig:phase-diagram}. Panel (a), (b), (c) and (d) represent
 respectively the multistability regions R4, R3, R1 and R2,
 respectively. The values of $\epsilon_1$ and $\epsilon_2$ are:
 $\epsilon_2=2.0$ and $\epsilon_1=2.21$ (panel (a)), $\epsilon_2=3.1$
 and $\epsilon_1=2.23$ (panel (b)), $\epsilon_2=2.5$ and
 $\epsilon_1=1.5$ (panel (c)), and $\epsilon_2=3.4$ and $\epsilon_1=2.5$
 (panel (d)). The data are obtained from numerical integration of the
 dynamics~(\ref{eq:eom}).}
 \label{fig:mul}
 \end{figure}

\begin{figure}[!ht]
	\hspace*{-1.2cm}
\includegraphics[width=10cm]{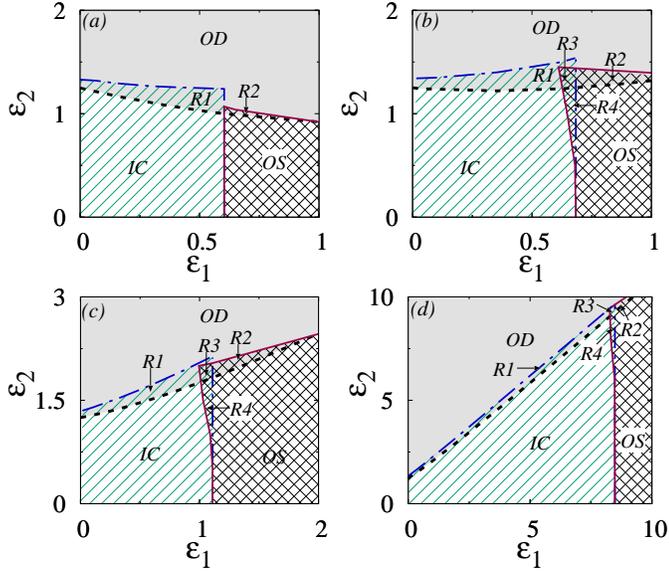}
\caption{For the model~(\ref{eq:eom}), the figure depicts the
two-parameter bifurcation diagram in the $(\epsilon_1,\epsilon_2)$-plane
for different $\alpha$ values given by $\alpha=0.1$ (panel (a)),
$\alpha=0.5$ (panel (b)), $\alpha=1.0$ (panel (c)) and $\alpha=1.5$
(panel (d)). The frequency distribution is the Lorentzian~(\ref{eq:lor})
with $\gamma=0.3$ and $\omega_0=0.6$. The regions of the different
states as well as the boundaries are obtained in the same manner
as in Fig.~\ref{fig:phase-diagram}. The regions R1, R2, R3, and R4
represent multistability or coexistence between IC-OD, between OS-OD,
between IC-OS-OD and between IC-OS, respectively.}
	\label{fig:1p}
\end{figure}

\begin{figure}[!]
%	\centering
\hspace*{-1cm}
	\includegraphics[width=9.5cm,height=4cm]{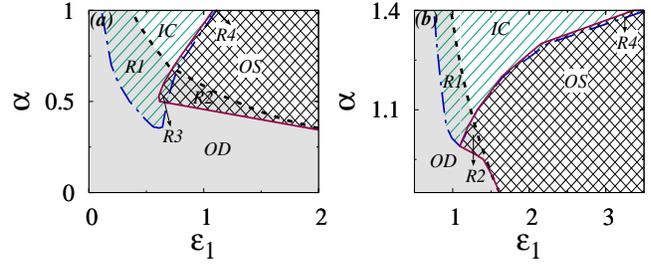}
	\caption{For the model~(\ref{eq:eom}) and considering for $\omega_j$'s the Lorentzian distribution~(\ref{eq:lor}) with
		$\gamma=0.3$ and $\omega_0=0.6$, the figure depicts the
		two-parameter bifurcation diagram in the
		($\alpha,\epsilon_1$)-plane for $\epsilon_2=1.4$ (panel (a)) and
                $\epsilon_2=2.0$ (panel (b)). We
		show here the various stable states and their boundaries, with bifurcation behavior
		observed as one crosses the
		different boundaries. The regions representing stable existence of the
		OD (Oscillation Death state), the IC (Incoherent 
		state) and the OS (Oscillatory Synchronized state) have been constructed
		by analysing the long-time numerical solution of the
		dynamics~(\ref{eq:eom}) for $N=10^4$, namely, by asking
		at given values of $\epsilon_1$ the quantity $r(t)$ at long times represents which one of the
		three possible behavior depicted in Fig.~\ref{fig:1}. The blue
		dot-dashed line, the black dashed line, and the maroon
		continuous 
		line are stability boundaries of the IC, the OD and the
		OS, respectively, and have been obtained by using the
		Ott-Antonsen-ansatz-based dynamics~(\ref{eq:r-dynamics}) and following
		the procedure of its analysis detailed in the text. The regions
		R1, R2, R3, and R4 represent multistability or
		coexistence between
		IC-OD, between OS-OD, between IC-OS-OD and between IC-OS, respectively. }

\label{fig:al}
\end{figure}
We now obtain the temporal behaviour of $r(t)$ for all the four
	multistability regions while starting from different initial states; the
	data are obtained from numerical integration of the
	dynamics~(\ref{eq:eom}).
	Figure~\ref{fig:mul}(a) shows that the dynamics initiated with either the IC or
	the OS remaining stabilized at these states (the inset shows that both these states
	remain stabilized at long times) when $\epsilon_1$ and $\epsilon_2$ have
	values in the R4
	region of the bifurcation diagram. Panel (b) shows that the dynamics
	initialized with either the IC or the OS or the OD remaining stabilized
	at these states when $\epsilon_1$ and $\epsilon_2$ have values in the R3
	region of the bifurcation diagram. Panels (c) and (d) represent in a
	similar manner the expected behavior in the multistability regions R1
	and R2, respectively.

We have until now analysed the bifurcation scenario in the
$(\epsilon_1,\epsilon_2)$-plane for $\alpha=1.3$. To illustrate the effect
of varying $\alpha$, we verified qualitatively similar bifurcation
diagrams for four different values of $\alpha$, namely, $\alpha=0.1,
~0.5,~1.0$ and $1.5$, shown in Fig.~\ref{fig:1p}, panels (a), (b), (c),
(d), respectively. The regions of the
                different states as well as the boundaries are
                obtained in the same manner as in 
                Fig.~\ref{fig:phase-diagram}. The
bifurcation diagram \ref{fig:1p}(a) for $\alpha=0.1$ is very similar to
the one for the dynamics~(\ref{eq:eom}) in the absence of asymmetry in
the interaction, i.e., with $\alpha=0$~\cite{Chandru:2020}. Even with a small
increase in $\alpha$ value to $\alpha=0.5$, we see the emergence of
regions R3 and R4, see panel (b). Moreover, we see that increase in
$\alpha$ values leads
to growth of the IC region.  In Fig.~\ref{fig:al}, we report for the dynamics~(\ref{eq:eom}) the
bifurcation diagram in $(\alpha,\epsilon_1)$-plane for different values of
$\epsilon_2$.  

%%%%%%%%%%%%%%%%%%%%%%%%%%%%%%%%%%%%%%%%%%%%%%%%%%%%%%%%%%%%%%%%%%%%%%%%%%%%%%%%%%%%%%%%%%%
\section{Conclusions}
\label{sec:conclusions}
In this work, we studied a nontrivial generalization of the Sakaguchi-Kuramoto model of spontaneous collective synchronization, by
considering an additional asymmetric interaction in the dynamics that
breaks the phase-shift symmetry of the model. We reported the emergence of a
very rich bifurcation diagram, including the existence of non-stationary
synchronized states arising from destruction of stationary
synchronized states. The bifurcation diagram shows existence of regions of
        two-state as well as three-state coexistence arising from asymmetric interaction in our model. Our results are based on numerical integration
of the dynamical equations as well as an exact analysis of the dynamics
by invoking the so-called Ott-Antonsen ansatz. It would be of immense
interest to explore how inclusion of inertial
effects~\cite{Gupta-inertia} in our model
modifies the bifurcation diagram~\ref{fig:phase-diagram}, an issue we are
working on at the moment.

%%%%%%%%%%%%%%%%%%%%%%%%%%%%%%%%%%%%%%%%%%%%%%%%%%%%%%%%%%%%%%%%%%%%%%%%%%%%%%%%%%%%%%%%%%%
%%%%%%%%%%%%%%%%%%%%%%%%%%%%%%%%%%%%%%%%%%%%%%%%%%%%%%%%%%%%%%%%%%%%%%%%%%%%%%%%%%%%%%%%%%%
\section*{Author's Contributions}
V.K.Chandrasekar and Shamik Gupta formulated the problem. M. Manoranjani carried out the simulations and calculations. All the authors discussed the results and drafted the manuscript.

\section*{Acknowledgements}
M.M. wishes to thank SASTRA Deemed University for research funds and for extending infrastructure support to carry out this work. S.G. acknowledges support from the Science
and Engineering Research Board (SERB), India under SERB-TARE scheme Grant No.
TAR/2018/000023, SERB-MATRICS scheme Grant No. MTR/2019/000560,  and SERB-CRG scheme Grant No. CRG/2020/000596.  He also thanks ICTP -- The Abdus Salam International Centre for Theoretical Physics,
Trieste, Italy for support under its Regular Associateship scheme. The work of V.K.C. is supported by the SERB-DST-MATRICS Grant No.
MTR/2018/000676 and  DST-CRG Project under Grant No. CRG/2020/004353.

\section*{Data Availability}
The data that support the findings of this study are available from the corresponding author upon reasonable request.

%%%%%%%%%%%%%%%%%%%%%%%%%%%%%%%%%%%%%%%%%%%%%%%%%%%%%%%%%%%%%%%%%%%%%%%%%%%%%%%%%%%%%%%%%%%

\end{document}